\documentclass[namedreferences]{solarphysics}
 \usepackage{graphicx}
 \usepackage{wrapfig}
\usepackage[hyperref,optionalrh]{spr-sola-addons}
\usepackage{spr-sola-addons}

\makeatletter
\let\jnl@style=\rmfamily
\def\ref@jnl#1{{\jnl@style#1}}%
\newcommand\aj{\ref@jnl{AJ}}%
\newcommand\actaa{\ref@jnl{Acta Astron.}}%
\newcommand\araa{\ref@jnl{ARA\&A}}%
\newcommand\apj{\ref@jnl{ApJ}}%
\newcommand\apjl{\ref@jnl{ApJ}}%
\newcommand\apjs{\ref@jnl{ApJS}}%
\newcommand\apss{\ref@jnl{Ap\&SS}}%
\newcommand\aap{\ref@jnl{A\&A}}%
\newcommand\aapr{\ref@jnl{A\&A~Rev.}}%
\newcommand\aaps{\ref@jnl{A\&AS}}%
\newcommand\azh{\ref@jnl{AZh}}%
\newcommand\baas{\ref@jnl{BAAS}}%
\newcommand\caa{\ref@jnl{Chinese Astron. Astrophys.}}%
\newcommand\cjaa{\ref@jnl{Chinese J. Astron. Astrophys.}}%
\newcommand\icarus{\ref@jnl{Icarus}}%
\newcommand\jcap{\ref@jnl{J. Cosmology Astropart. Phys.}}%
\newcommand\jrasc{\ref@jnl{JRASC}}%
\newcommand\memras{\ref@jnl{MmRAS}}%
\newcommand\mnras{\ref@jnl{MNRAS}}%
\newcommand\na{\ref@jnl{New A}}%
\newcommand\nar{\ref@jnl{New A Rev.}}%
\newcommand\pra{\ref@jnl{Phys.~Rev.~A}}%
\newcommand\prb{\ref@jnl{Phys.~Rev.~B}}%
\newcommand\prc{\ref@jnl{Phys.~Rev.~C}}%
\newcommand\prd{\ref@jnl{Phys.~Rev.~D}}%
\newcommand\pre{\ref@jnl{Phys.~Rev.~E}}%
\newcommand\prl{\ref@jnl{Phys.~Rev.~Lett.}}%
\newcommand\pasa{\ref@jnl{PASA}}%
\newcommand\pasp{\ref@jnl{PASP}}%
\newcommand\pasj{\ref@jnl{PASJ}}%
\newcommand\qjras{\ref@jnl{QJRAS}}%
\newcommand\rmxaa{\ref@jnl{Rev. Mexicana Astron. Astrofis.}}%
\newcommand\rnaas{\ref@jnl{Res.~Notes~AAS}}%
\newcommand\skytel{\ref@jnl{S\&T}}%
\newcommand\solphys{\ref@jnl{Sol.~Phys.}}%
\newcommand\sovast{\ref@jnl{Soviet~Ast.}}%
\newcommand\ssr{\ref@jnl{Space~Sci.~Rev.}}%
\newcommand\zap{\ref@jnl{ZAp}}%
\newcommand\nat{\ref@jnl{Nature}}%
\newcommand\iaucirc{\ref@jnl{IAU~Circ.}}%
\newcommand\aplett{\ref@jnl{Astrophys.~Lett.}}%
\newcommand\apspr{\ref@jnl{Astrophys.~Space~Phys.~Res.}}%
\newcommand\bain{\ref@jnl{Bull.~Astron.~Inst.~Netherlands}}%
\newcommand\fcp{\ref@jnl{Fund.~Cosmic~Phys.}}%
\newcommand\gca{\ref@jnl{Geochim.~Cosmochim.~Acta}}%
\newcommand\grl{\ref@jnl{Geophys.~Res.~Lett.}}%
\newcommand\jcp{\ref@jnl{J.~Chem.~Phys.}}%
\newcommand\jgr{\ref@jnl{J.~Geophys.~Res.}}%
\newcommand\jqsrt{\ref@jnl{J.~Quant.~Spec.~Radiat.~Transf.}}%
\newcommand\memsai{\ref@jnl{Mem.~Soc.~Astron.~Italiana}}%
\newcommand\nphysa{\ref@jnl{Nucl.~Phys.~A}}%
\newcommand\physrep{\ref@jnl{Phys.~Rep.}}%
\newcommand\physscr{\ref@jnl{Phys.~Scr}}%
\newcommand\planss{\ref@jnl{Planet.~Space~Sci.}}%
\newcommand\procspie{\ref@jnl{Proc.~SPIE}}%

\renewcommand*{\@biblabel}[1]{\hfill}
\renewcommand*{\@cite}[1]{#1}
\renewcommand\section{\@startsection {section}{1}{\z@}%
                                     {-3.5ex \@plus -1ex \@minus -.2ex}%
                                     {2.3ex \@plus .2ex}%
                                     {\normalfont\large\bfseries}}
\renewcommand\subsection{\@startsection {subsection}{2}{\z@}%
                                        {-3.5ex \@plus -1ex \@minus -0.2ex}%
                                        {1.5ex \@plus 0.2ex}%
                                        {\normalfont\normalsize\bfseries}}
\renewcommand\subsubsection{\@startsection {subsubsection}{3}{\z@}%
                                        {-2.0ex \@plus -0.8ex \@minus -0.2ex}%
                                        {1.0ex \@plus 0.2ex}%
                                        {\normalfont\normalsize\itshape}}
\renewcommand\paragraph{\@startsection {paragraph}{4}{\z@}%
                                        {2.0ex \@plus0.8ex \@minus 0.2ex}%
                                        {-1em}%
                                        {\normalfont\normalsize\itshape}}
\makeatother

\def\bul{\raise.2ex\hbox{$\bullet$}}
\def\solar{\odot}

\begin{document}

\begin{article}
\begin{opening}

\title{Photospheric Velocities Measured at Mt. Wilson Show Zonal and Sectoral Flows Compose the Torsional Oscillations}

\author[addressref={aff1},corref,email={ulrich@astro.ucla.edu}]{\inits{R. K.}\fnm{Roger K.}~\lnm{Ulrich}\orcid{0000-0003-4041-4657}}

\author[addressref={aff1},email={tran@astro.ucla.edu}]{\fnm{Tham}~\lnm{Tran}\orcid{0000-0003-1924-1763}}

\author[addressref={aff1}]{\inits{E.}~\fnm{John}~\lnm{Boyden}~}

\address[id=aff1] {Dept.\ of Phys.\ and Astro., 
University of Calif.\ at Los Angeles 
475 Portola Plaza, Los Angeles, CA 90095-1547
}



\runningauthor{Ulrich, Tran and Boyden}
\runningtitle{Zonal and Sectoral Flows}

\begin{abstract}
The methods for reducing the observations from the 150-foot Tower Telescope on Mt.~Wilson are reviewed and a new method for determining the North/South (sectoral) and the East/West (zonal) velocity components is described
and applied. Due to a calibration problem with the data prior to 1983,
only observations between 1983 and 2013 are presented at this time. After subtraction
of latitude dependent averages over the 30-year period of
 observation the residual deviations in the sectoral and zonal flow velocities are well synchronized and correspond to what is widely recognized as the Torsional Oscillations.  Both flow components need to be included in any model that replicates the Torsional Oscillations.
\end{abstract}
\keywords{Helioseismology, Observations; Velocity Fields, Interior; Velocity Fields, Photosphere; Supergranulation }

\end{opening}
\section{Introduction}
\label{Intro}
This paper reports on progress in the analysis of the archival data base from the observations made at the 150-foot Tower Telescope at the Mt.\ Wilson Observatory (MWO).  A new method of determining large scale flow velocities and their variations is presented here along with results showing the close association between surface velocities deviations going E/W and N/S.  These new results will help in the study of 
the nature of the solar dynamo which is currently under intense study using complementary helioseismic and feature tracking methods.  Those methods also yield E/W and N/S transverse velocities that apply to a range of solar interior depths.  Small time dependent variations derived from these velocities are of particular interest.  

The totality of the collected magnetogram/Dopplergram observations is a resource that allows the study of temporal variation in the solar rotation rate (\cite{1976ApJ...210L.159H,1983SoPh...83..321H,1984SoPh...93..171G,1984ARA&A..22..131H,1984ApJ...283..385G}).  An important finding of these studies was the identification
of an E/W moving pattern of velocity called the Torsional Oscillations by \cite{1980ApJ...239L..33H,1982SoPh...75..161L,1982SoPh...80..373L} and \cite{1983SoPh...82..437H}. 
Proposals by \cite{1981A&A....94L..17S,1981ApJ...247.1102Y} suggested the Torsional Oscillations are produced by Lorenz forces from the solar cycle magnetic fields.  The work by \cite{1983SoPh...82..437H}, however, pointed out that the Torsional Oscillations occur at times and places where there are no strong magnetic fields. 

The Torsional Oscillations have been confirmed through Helioseismology observations (\cite{1997ApJ...482L.207K,1999ApJ...523L.181S,2000ApJ...533L.163H,2021ApJ...908L..50G}); (see the review provided by \cite{2009LRSP....6....1H}).  Global dynamo modeling by \cite{2016ApJ...828L...3G} finds that magnetic tension at the tachocline lower boundary of the convection zone can generate a dynamo wave that propagates to the solar surface.  Subsequently, \cite{2019ApJ...871L..20K} found support for this idea from a Principal Component Analysis (\cite{2013PCA}) of the zonal acceleration derived from global helioseismology.     

The analysis of the present paper raises additional questions - a) ``What is the direction of the flow?" and b) ``What is the reference function that we should subtract?''  The time scale for changes in the flow is an important part of characterizing the motion.  Since we do not know the cause for the dynamics we prefer to use the terms ``zonal flow" for the E/W motions and ``sectoral flow" for the N/S motions.   For the Torsional Oscillations a differential rotation function is subtracted.  For the sectoral flow there is a process called meridional circulation which is a consequence of stellar rotation [see for example Sect.\ 42 of \cite{1990sse..book.....K}] which generates a steady flow with upwelling at the equator, surface flow from equator to pole, downdraft at the pole and a return flow through the solar envelope, not necessarily at the convective/radiative boundary. Since the total solar angular momentum is not variable, this flow should also not be variable.  Assuming the flow variations we find are related to the solar cycle, we need to subtract a similar background circulation function which we will refer to as the meridional circulation function.  We find both the  differential rotation function and the meridional circulation function by averaging over the full time span of the record we are using.  Data outside our time frame could alter these functions which would alter the variations we determine.  We will show in this paper that the zonal and sectoral pair of flows have variations which are both part of the Torsional Oscillations.

The full nature of the Torsional Oscillations has not been evident because only the east-west
component of the velocity vector was part of the variation. Addition of north-south velocity component as part of the process may help identify the underlying cause.
Although surface meridional circulation due to rotation generally originates in the equatorial zone and moves toward the poles, the work by \cite{2010ApJ...725..658U}, especially figure 6, showed that there are periods when the flow is toward the equator.  That work also showed that there are strong year-to-year changes.    

The study by \cite{2018SoPh..293..145K} found some evidence of equator-ward motion of the sectoral flow.  More extensive studies by \cite{2018SoPh..293..145K,2020SoPh..295...47K,2021ApJ...908L..50G} have found variations in the sectoral flow in bands that migrate toward the solar equator during the solar cycle.  They have correlated the regions on the sun with magnetic activity with the strength of these sectoral flows.  There is a general confirmation of the model by \cite{2003SoPh..213....1S} wherein excess emission in magnetic regions causes cooling and sinking of the matter in magnetized area.  Studies by \cite{2020SoPh..295...47K,2021ApJ...908L..50G} have found evidence in support of this model but there remains the difficulty that the motions are present before there are strong magnetic fields.  Our new method of analysis shows that the zonal and sectoral components are closely related throughout the solar cycle.

\section{Early Record -- Why Start in 1983}

The digital data record of the 150-foot tower telescope at the Mt.\ Wilson Observatory begins in 1967 and our project aims to utilize this full record.  Solar differential rotation rates have typically been represented by a three term series in $\sin(\phi)$ where $\phi$ is the latitude and the lead term is $A$. Our equation (\ref{rotrate}) below is typical.
However, the calibration of the $A$ coefficient was flawed in the years prior to 1983.  The calibration of the early observations came from a displacement of the entrance slit a known distance so that a known signal was introduced.  This approach which was introduced by \cite{1953ApJ...118..387B} as part of the magnetograph design involved the controlled displacement of the spectral line from its centered position and the measurement of the resulting artificial signal.  Because of variations in the size of the displacement, the calculated calibration coefficients were not stable so that the determined $A$ rotation rates and the calibration coefficients are inversely related. While a correction to the $A$ values can be derived by assuming the calibration coefficient is fixed, the earlier data will remain less reliable and we have concentrated on the data obtained after 1983. 

\section{Initial Reductions of Surface Velocities}
\label{InitReduct}
\subsection{Background}
The instrument and measuring system setup and calibration for the Mt.\ Wilson Magnetograph is well described by \cite{1983SoPh...87..195H}\ and details of the setup and operation of the system can be found in that reference.  We review some of these details because they influence results we describe here.

The data from the 150-foot tower project applies to the solar surface yielding magnetic field $B$, Doppler velocity $V$ and line intensity $I$ as functions of heliographic latitude $\phi$, central meridian angle $\delta L$ which is the westward angle from the central meridian to the pixel and time $t$.  An important property of this dataset from the 150-foot tower is its early start date.  A complication dealt with in the reduction of this data comes from the fact that observed data is not on a regular grid of spatial position and the temporal spacing is irregular both within individual observations and between successive observations.  Consequently interpolations are required to produce tables of the observables in regularly spaced grids.

The polarization analysis uses a rapidly oscillating K*DP crystal combined with a Glan-Thompson prism to alternate the state of circular polarization allowed through the spectrograph.  The spectral sampling is done by rectangular shaped fiber-optic light pipes whose long dimension entrance windows are perpendicular to the dispersion direction.  Paired samples of the spectrum from the red, $R$, and blue, $B$, wings of the sampled line are sent to two pulse-counting photomultiplier tubes yielding count values stored in four registers:
 $R^+,R^-,B^+,B^-$ where the superscript indicates the polarization state.  At regular intervals the sum values are recorded on disk and the registers are cleared.  Each data record includes the four sample values, the time along with the location of the servoed stage carrying the sampling rectangles and the position of the servos controlling the position of the solar image. These packets of data are recorded sequentially on a magnetic disk.  At the end of the observation these records get transferred to magnetic tape.  These files have a name structure where the letter 'M' is followed by YYYYMMDD\_HHHH\_MM.01 where YYYY is the year, MM is the month, DD is the day and HHHH\_MM is the time.  We refer to these as the 01 files.

While the 01 files include all the necessary information, they are in a difficult-to-use format - each data element has a different time, the spacing between the scan lines is not constant and the elements are not lined up vertically. We address these problem in steps:  1) Each file is converted to an internal format we call an {\it MVI} file which has three images which are shifted by differential rotation to be at the same time and to compensate for the guiding system which is imperfect.  These shifts produce a grid which is rectangular. These images show the magnetic, velocity and intensity variables, 2) For each day the pixels of the {\it MVI} images are shifted at constant latitude using the local rotation rate to a fixed time and are averaged together, 3) A set of {\it fits} files is created for each average and a text file called {\it VMREC} is written.  The {\it VMREC} file is derived by appling geometric corrections to the background limb shift function [see section (\ref{NewRate})], along with first guess differential rotation and meridional circulation laws to yield position dependent line-of-sight velocity deviation and magnetic field variables as functions of $x,y$ indicies $i$, $j$.  Steps to derive an update to the flow functions from the {\it VMREC} file are described in the following sections.

\subsection{Differential Rotation and Sectoral Flow}
\label{dvmerid}
A primary output of the observing program at the 150-foot tower telescope has been the study of the time dependence of the solar surface velocities. The system provides Doppler shifts of the spectral line which are converted to rotation rate and N/S flow velocity using the spherical trigonometry algorithms from \cite{1970SoPh...12...23H}.  In addition we modify the sign for the N/S velocity by reversing the sign in the southern hemisphere so a positive value represents flow toward the poles.  The largest feature of these velocities is the differential rotation which has a large East/West component.  Flow variability with small amplitude cannot be detected superposed on the differential rotation unless the main pattern is subtracted.  This has been done using a polynomial with the coefficients $(A,B,C)$ in the representation
\begin{equation} 
\Omega(\phi,t) = A(t)+B(t)\sin^2(\phi)+C(t)\sin^4(\phi) \; \mu\,{\rm rad/s}  \label{rotrate}
\end{equation}
where $\Omega$ is the rotation rate and $\phi$ is the heliographic latitude.  For many years this polynomial was fitted to each observation.  This representation is imprecise with there being no expectation that the flow should have this structure.  To address this concern and still allow us to detect small and variable structures we have run the 01 files through a new initial reduction with the modification that the $B,C$ pair of coefficients is held fixed.  This produces a new archive of {\it MVI} files and other products from which we can retrieve the observed flows but which has a static, first guess differential rotation offset curve subtracted.  The curve we have used is:
\begin{equation}
\Delta\Omega_0(\phi)=-0.409999*[\sin^2(\phi) + 1.0216295*\sin^4(\phi)]\; \mu\,{\rm rad/s}\ . \label{Drotrate0}
\end{equation}
Note that the subtracted quantity is negative so the rotation rate is increased to become more nearly constant with latitude.
The flow remaining after equation (\ref{Drotrate0}) has been applied can be described by either a rotation velocity deviation $\delta V_{\rm rot}$ or as an angular rotation rate deviation $\delta\omega$.
The conversion from angular rotation rate deviation to rotation velocity deviation is:
\begin{equation}
\delta V_{\rm Rot}(\phi,t)=\delta\omega(\phi,t)*R_\solar \cos(\phi)\ .  \label{dvelocity}
\end{equation}
We use equation (\ref{dvelocity}) to convert rotation rates or rotation rate deviations to rotation velocities or rotation velocity deviations and vice versa as needed.


We observe the line-of-sight velocity $V_{\rm los}$ at each point. Using the geometry as described by \cite{1970SoPh...12...23H} we convert this velocity to a rotation rate $\Omega(\phi,t)$ and a rotation rate deviation:
\begin{equation} 
\delta\omega_0(\phi,t) = \Omega(\phi,t)-\Delta\Omega_0(\phi)\ .
\end{equation}
The values of $\delta\omega_0(\phi,t)$ along with magnetic field strength and line intensity can be read into standard {\it fits} files which we pick to be daily averages with dimension $256\times256$ in $\delta L$ and $\phi$.

 The differential rotation rate offset from Equation (\ref{Drotrate0}) was adopted in 1987 and those coefficients
do not accurately represent the rotation rate over the full time series.  Consequently, we work with a reference rotation rate equation (\ref{Drotratref}) and we get rotation rate deviations relative to the actual 30-year average equation (\ref{drotrate})
\begin{equation}
\delta\omega_{\rm ref}(\phi) =  <\Omega(\phi,t)>_{\rm average}-\Delta\Omega_0(\phi)  \label{Drotratref}
\end{equation}
\begin{equation}
\delta\omega(\phi,t) = \delta\omega_0(\phi,t) - \delta\omega_{\rm ref}(\phi)\ .  \label{drotrate}
\end{equation} 
In order to recover the rotation rates themselves we have to use the sum 
\begin{equation}
\Omega(\phi,t) = \delta\omega(\phi,t)+\delta\omega_{\rm ref}(\phi)+\Delta\Omega_0(\phi)\ .\label{newrotrate}
\end{equation} 

The original program reduction yielded the first guess angular rotation rate as in equation (\ref{rotrate}) whereas our new reduction yields local line-of-sight velocity deviations relative to the projection of the 30-year average rotation rate.  
The new rotation velocity deviation is then
\begin{equation}
\delta V_{\rm rot}={d\delta V_{\rm los}\over d\sin(\delta L)}\ .\label{Vrot}
\end{equation} 
This equation provides a value of $\delta V_{\rm rot}$ for each pixel of the image.

The first guess background meridional flow law adopted at the same time as the differential rotation and is:
\begin{equation}
V_{\rm mer}(\phi,t) = [\cos({\delta L})\sin(\phi)\cos(B_0)-\cos(\phi)\sin(B_0)]\{G\sin(\phi)+H\sin(2\phi)\}    \label{dvmerid}
\end{equation} 
where $G=-28.692\,{\rm m\,s}^{-1}$ and $H=35.240\,{\rm m\,s}^{-1}$.  The sectoral velocity deviation we seek, $\delta V_{\rm sec}$, is relative to this function corrected the same way as $\delta V_{\rm rot}$.
This function was subtracted during the initial reduction and has to be restored now as part of the determination of the 30-year average.

\section{New Rotation Rate Algorithm}

\label{NewRate}

The method we used previously starts with the initial reduction with the algorithm described by \cite{1983SoPh...87..195H} which uses pixels from most of the solar disk including those near the limb.  The intensity gradient approaching the limb causes scattered light to have a significant effect on the determined rotation law.  The amount of scattered light in each solar image is determined by recording intensity values for pixels near but outside the limb.  The average from such pixels is used to quantify the scattered light which after correlation with the measured rotation rate allows the derivation of a correction to the rotation rate.  For observations with high scattered light, the correction can be large.  Our new method avoids the need for this correction.

The scattered light correction is important near the limb due to the steep gradient in intensity as the limb is approached.  The position of each pixel is defined by $\phi$, and $\delta L$. The new method avoids the steep gradient zone by requiring $|\sin(\delta L)|<0.85$ which leaves out the near-limb pixels.  The heliographic geometry and resolution of the observed Doppler velocities into rotation and meridional circulation was described by \cite{1970SoPh...12...23H} and \cite{2006A&A...449..791T}.  

Our task of determining the zonal and sectoral flow rates is complicated by another effect known as the limb shift.  Due to correlations between vertical velocity and spectral line intensity due to temperature changes, photospheric spectral lines are shifted as a function of the angle the emergent ray makes relative to the local vertical.   Both the flow fields and the limb shift contribute to the Doppler shift velocity, $V_{\rm los}$,  of each pixel and we wish to isolate only the material flow.  We find the limb shift by using an equatorial band 4$^\circ$ wide out of the velocity array before the subtraction of the differential rotation.  These entries are minimally influenced by the sectoral flow and symmetrically influenced by the zonal flow so we can use this band to derive the limb shift function. The relationship between the limb shift and the zonal and sectoral material flows was discussed in detail by \cite{1988SoPh..117..291U}. 

As described in Section(\ref{InitReduct}) the scanning system at the 150-foot tower at Mt.\ Wilson produces a data file {\it VMREC} which gives position indices $i$ and $j$ along with daily averages of the line-of-sight component of the rotation velocity deviations, and line-of-sight magnetic field strength in $256\times256$ arrays. Our working quantities are the line-of-sight component of the rotation velocity deviations $\delta V_{\rm los}$ and the magnetic fields.  We calculate these for the latitude bands 4$^\circ$ wide defined below using our new method.

\begin{figure}
\resizebox{4.5in}{!}{\includegraphics{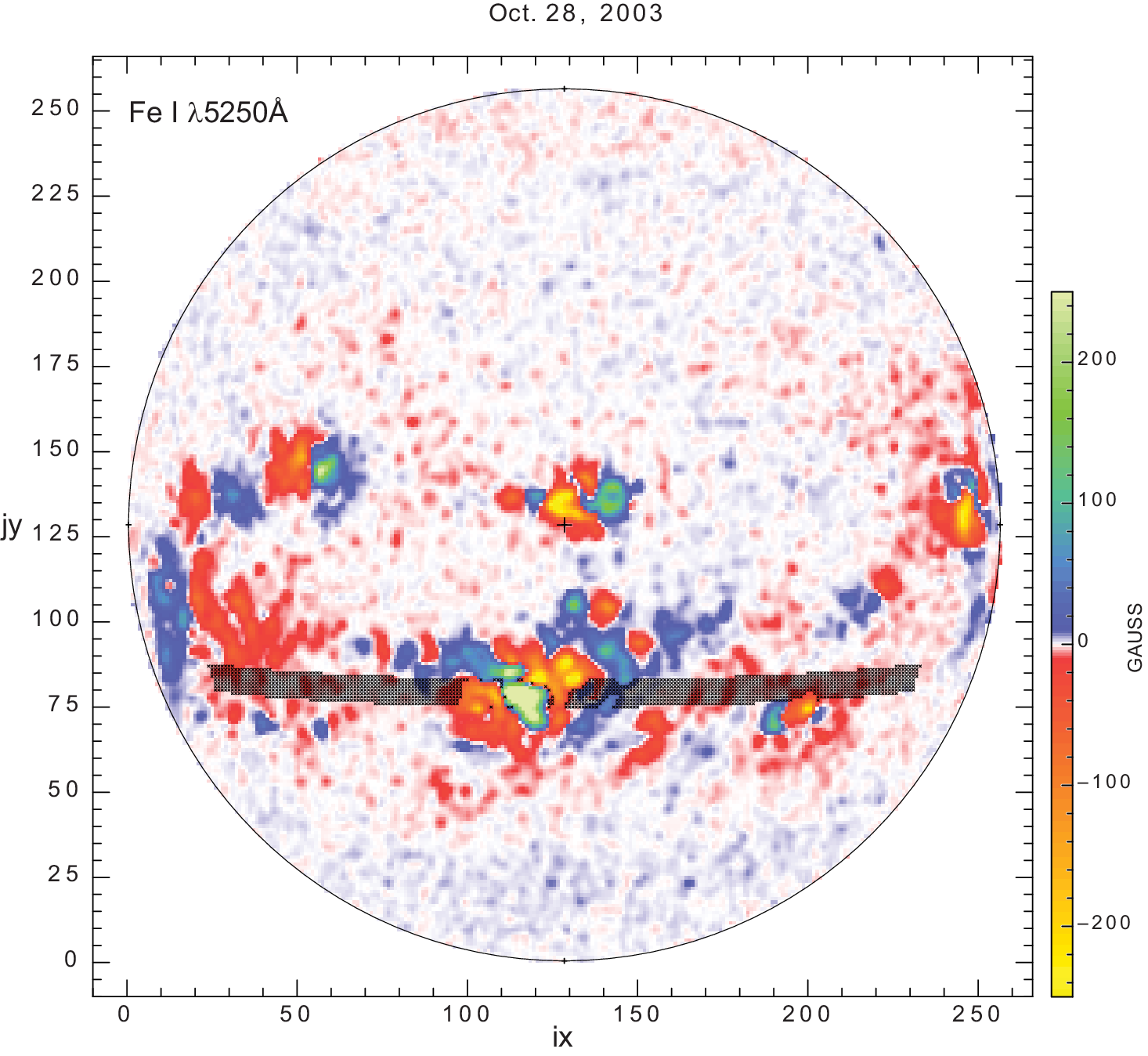}}
\caption{This figure shows a sample daily average magnetogram from Oct. 28, 2003.  The average is formed by spatial interpolating to a rectangular of dimension $256\times256$ in $\delta L$ and $\phi$ and temporal shifting of each pixel using differential rotation to where the point would be at 20.0 UT.   The pixel positions marked out with an $X$ are those which were included in the algorithm for determining the rotation rate.  These pixels are in latitude band number 16 which covers the range $20^\circ S<\phi<16^\circ S$.  In addition accepted pixels have $|\sin(\delta L)|<0.85$ and have an absolute magnetic field strength $|B|$ less than 30G.  Notice that along the lines where the magnetic field changes sign there are a number of included pixels because the field direction is nearly horizontal.  Although this figure  shows the magnetogram, the velocities for the shown pixels are extracted at the corresponding points on the Dopplergram.
}
\label{figone}
\end{figure}

\begin{figure}
\resizebox{4.5in}{!}{\includegraphics{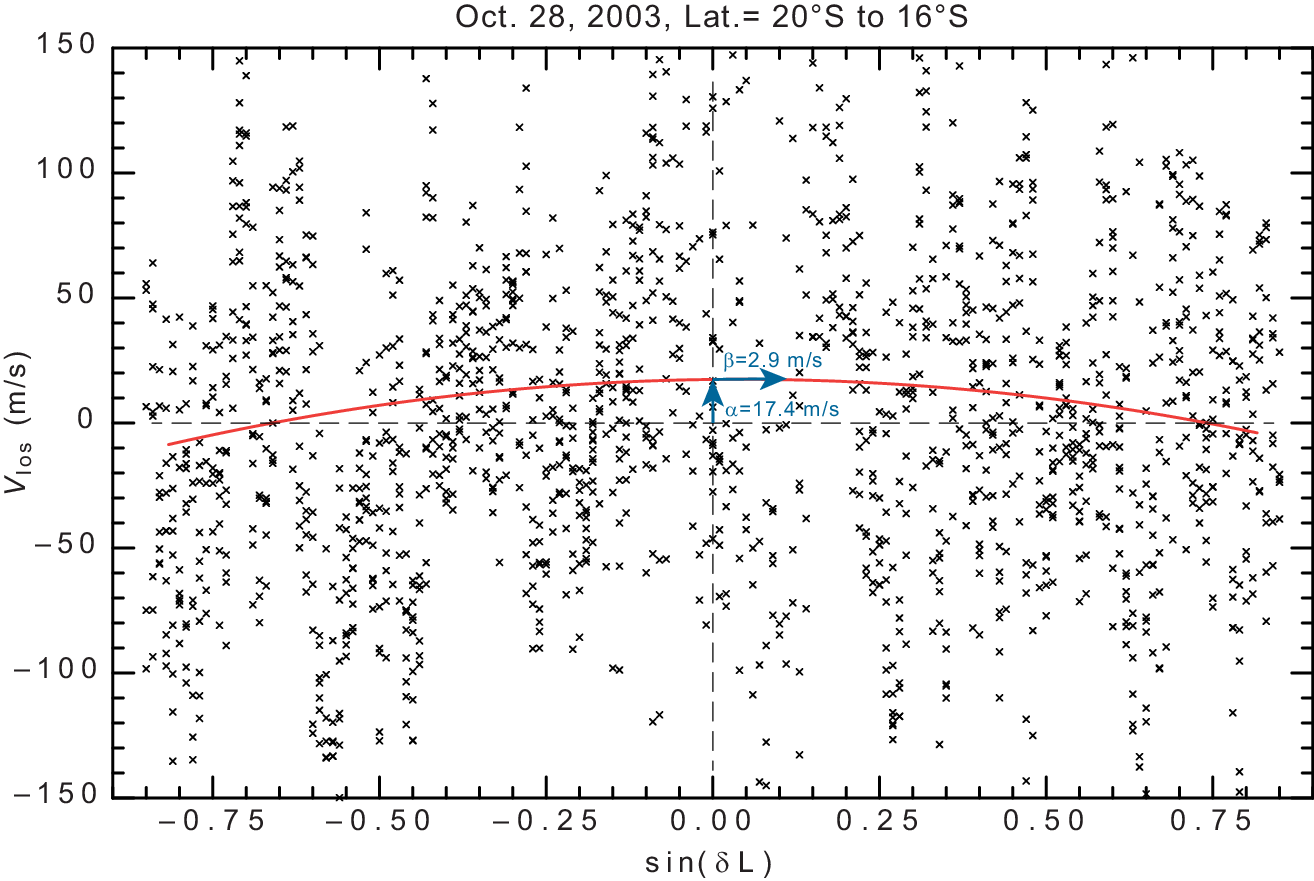}}
\caption{This is the sample latitude band shown in Figure(\ref{figone}).  Each selected pixel is shown with an $X$ symbol at its line-of-sight velocity $V_{\rm los}$ and  sine central meridian angle: $\sin(\delta L)$.  The red line is the quadratic fit to those pixels and the fitting coefficients $\alpha$ and $\beta$ are shown by the blue arrows.  The value of $\beta$ is indicated by the very small but non-zero slope of the arrow.  The rms variation of both $\alpha$ and $\beta$ is about 15 m/s overall and the value of $\beta$ is unusually small in this case.  This figure is for a single day of observation while the quantities used for measurement of the surface flows are averages over six Carrington rotations.}
\label{figtwo}
\end{figure}

The new method begins by using the latitude $\phi$ to define bands which are each $4^\circ$ wide:
\[j_{\rm band} = {\rm Round}((\phi+82)/4)\] 
and
collects pixels within each band.  We refer to $j_{\rm band}$ as the band number.
The selected pixels also have $|\sin(\delta L)|<0.85$.  The Doppler shift velocities for each daily observation and each latitude band is
then fit by a quadratic function of $\sin(\delta L)$ yielding three coefficients $(\alpha,\beta,\gamma)$:
\begin{equation}
V_{los}(\delta L)=\alpha+\beta\sin(\delta L)+\gamma\sin^2(\delta L)\ .\label{quadfit}
\end{equation} 
The coefficients are: $\alpha$ is the daily offset velocity due to the sectoral flow, $\beta$ gives the daily rotation velocity deviation and $\gamma$ allows the flexibility to match the other two coefficients. The rotation rate deviation, $\delta V_{\rm rot}(\phi,t)$, is then found from the average of the $\beta$ values over 6 Carrington rotations. 

Each daily result is influenced by supergranulation noise.  The only way to reduce this noise is through averaging.  For that reason we combine the daily results for $\delta V_{\rm rot}$ and $\delta V_{\rm sec}$ in 6 or 12 Carrington rotation averages.  When we use 6 CR averages, we apply an even-odd filter, (0.25,0.50,0.25), to the time series to remove the effect of polar tilt from the final series.

As an example we show in figure (\ref{figtwo}) the $\sin(\delta L)$ dependent points from Oct.\ 28, 2003 with each point value and the fit to equation (\ref{quadfit}). This is the same day and latitude band shown in Figure (\ref{figone}).  This particular day corresponds to Carrington Rotation 2009.1936 and year 2003.8222.  The fit coefficients for latitude band 16 for this day are $\alpha=17.431, \beta=2.874 {\rm\ and\ } \gamma= -71.142$ all with units of m/s since the $\sin(\delta L)$ factor is dimensionless.

\section{Data Reduction Steps}
\label{RedSteps}
The initial step carries out the latitude band fit described in the preceding section for the 7939 days of observation in the record.  Each day produces a line of output with 42 entries of $\alpha$ and $\beta$.  These are combined into a Microsoft Excel file which is output into two text files, one for the Northern hemisphere and one for the Southern hemisphere.  Entries in these files are Null if fewer than 8 points are in the latitude band or if the slope is greater than 100 m/s.

\begin{table}
\caption{This table gives the angular rotation rate from Equation (\ref{newrotrate}) as a function of latitude band averaged over the 30 Years of this data set.  The error of the mean, $\delta\Omega$ EoM, is also given based on the variance of the data values and the number of points included.
\label{tableone}
}
\begin{tabular}{c}
\resizebox{3.5in}{!}{\includegraphics{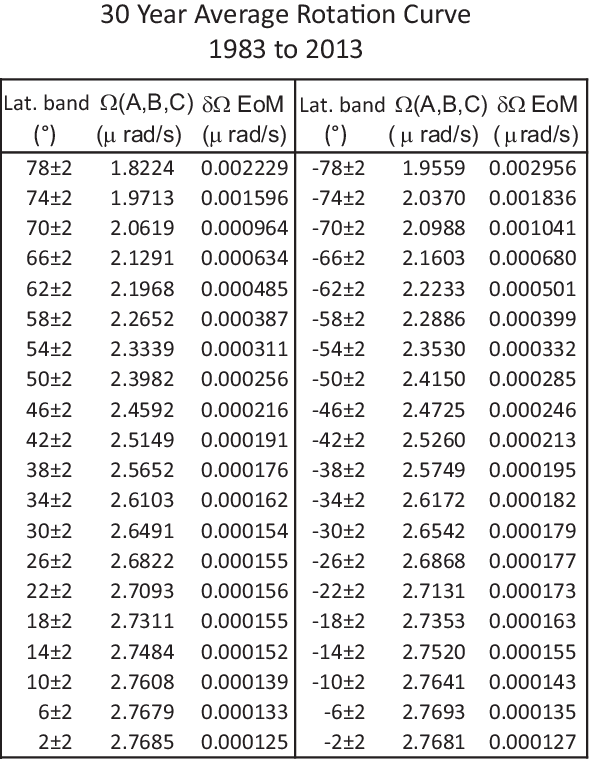}
}
\end{tabular}
\end{table}

The time series for each latitude band is cleaned independently using a procedure based on the variance.  The variance for each group of 120 points is calculated and a quadratic function of time is fitted to them.  The point with the largest absolute deviation from the fit is found and removed and the variance is recalculated.  If this variance has decreased by more than 8\%, the point is replaced by the fitted value and next largest deviation is found and removed. The test is repeated until the variance changes by less than 8\%. The last deleted point is then restored and the analysis moves to the next group of 120 points.  For each latitude band's time sequence we have about 50 points out of the 7939 that are removed this way.  The points removed, however, would have a strongly adverse effect on the derived velocities if left in. The highest latitude sequences have fewer observations due to the requirement that the velocities stay below 100 m/s or due to the failure to have more than 8 observations at each accepted latitude and time.  The same cleaning process is applied to the sectoral flow velocities.

\begin{figure}
\begin{center}
\resizebox{4.5in}{!}{\includegraphics{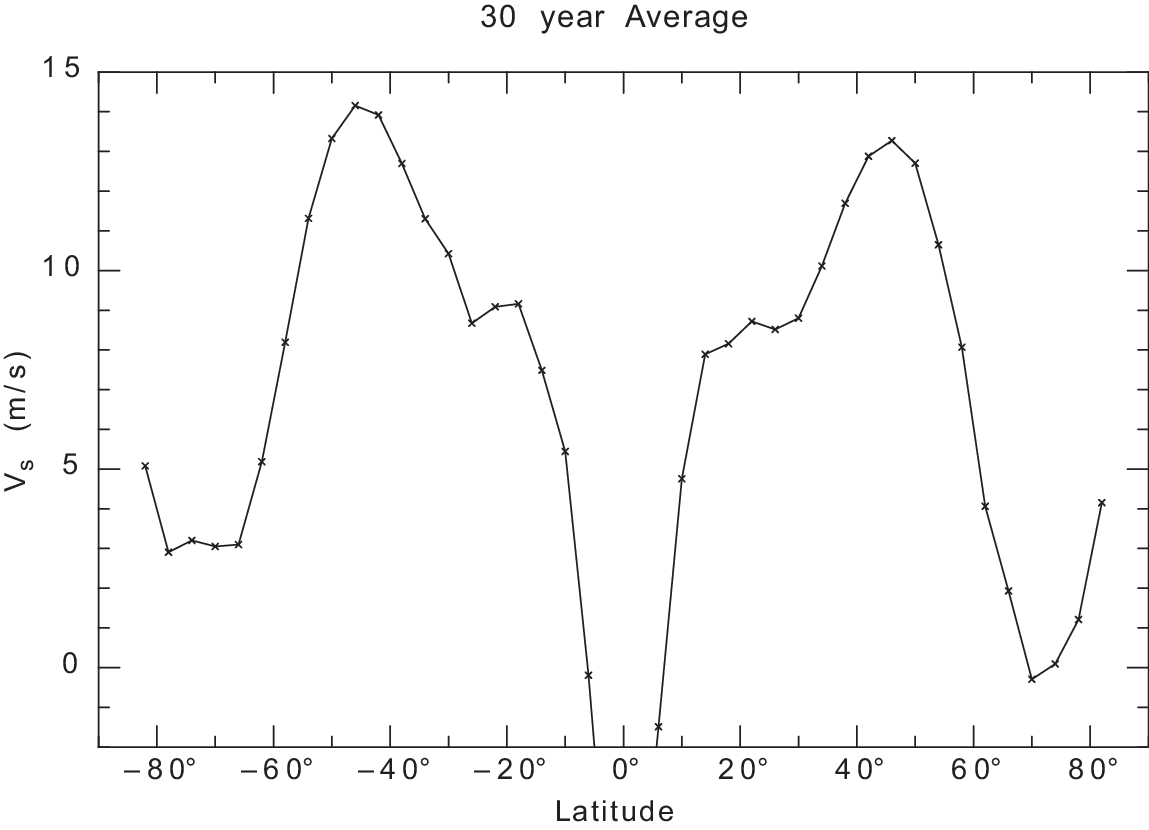}}
\caption{This figure shows the 30-year average of the sectoral velocity as a function of latitude. This average is the combination of the offset during the initial reduction and the average of the individual offsets determined with this new method. See the discussion of the sign convention in the text of section(\ref{RedSteps}). }
\label{figfour}
\end{center}
\end{figure}

Additionally the rotation rate as a function of time and latitude is calculated by dividing the rotation velocities by $R_\solar \cos(\phi)$.  Next we compute the 30-year averages of the rotation velocities, the rotation rates and the sectoral flow velocities.
The 30-year averages of the rotation rates are given in Table \ref{tableone}.  The North/South difference in this table is likely of solar origin.  The sectoral velocities have been treated similarly with the exception that we have reversed the sign in the southern hemisphere.  For coordinates on a sphere positive velocity values flow from south toward north.  For the sectoral velocities we desire the positive velocities to be flowing away from the equator and toward the poles.  For the southern hemisphere this requires a sign reversal.  In addition the initial reductions subtracted the final term with $G$ and $H$ 
in Equation (10) of \cite{1988SoPh..117..291U}\ as indicated above in Equation (\ref{dvmerid}).  To get the average sectoral flow velocity we have to combine this term and the average offset of the arrays in the archive.  This sum including the sign reversal is shown in Figure (\ref{figfour}).

\section{Rotational and Meridional Flow Maps}

\begin{figure}
\begin{center}
\resizebox{4.5in}{!}{\includegraphics{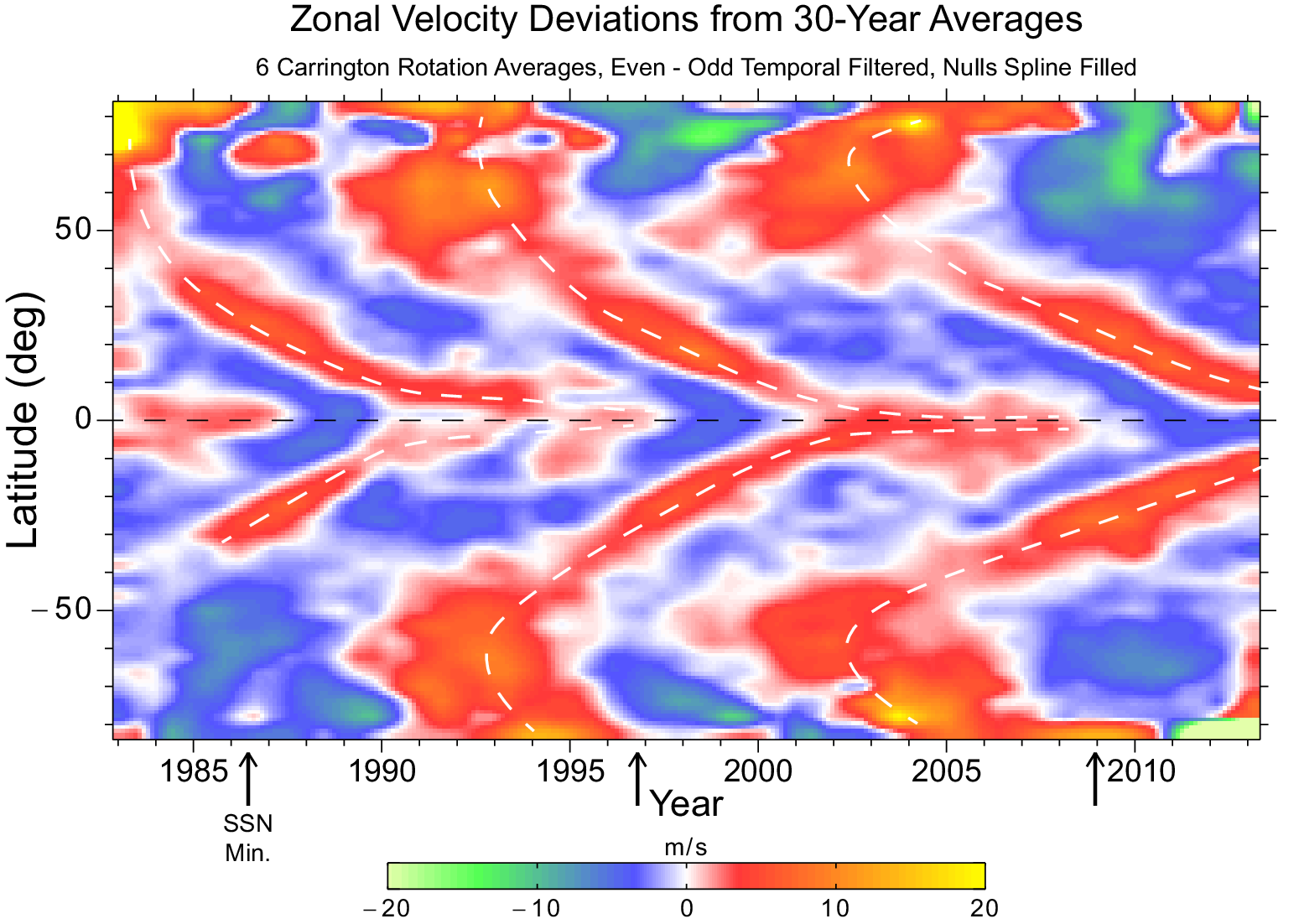}}
\resizebox{4.5in}{!}{\includegraphics{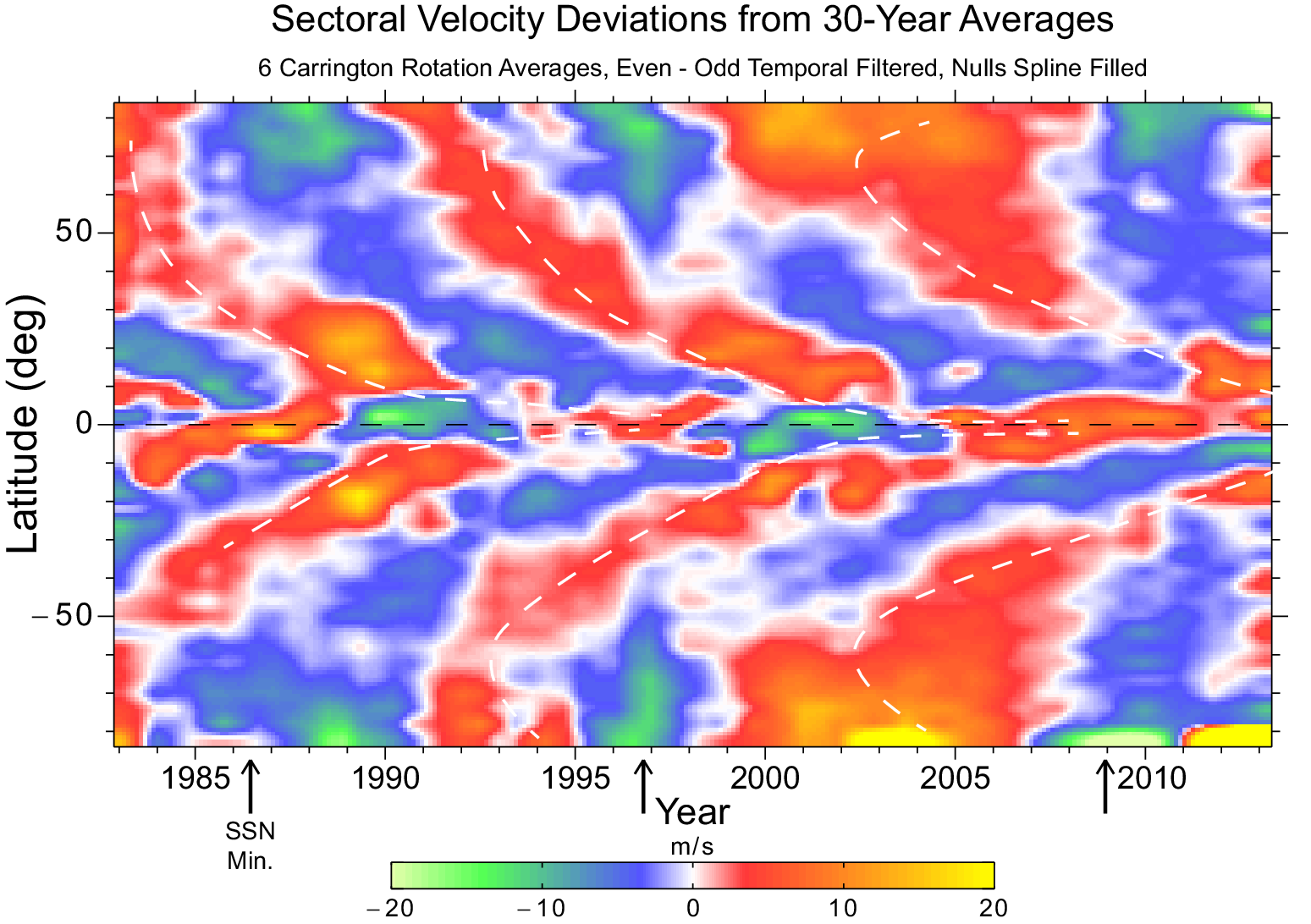}}
\caption{The velocity deviations shown in these two figures have been derived according to the algorithms described in the previous sections. Temporal filtering with a (0.25,0.5,0.25) even/odd filter was applied to both maps prior to plotting. The two components - East West in the zonal plot and North South in the sectoral plot together constitute the process known as the Torsional Oscillations.  The dashed white line traces the latitude of the maximum speedup at each time on the zonal map.  These line locations are transfered to the same latitudes in the sectoral plot map.  For latitudes near 10$^\circ$, the time of maximum poleward flow speed in this second map is a year or so after the position of this line, but for most of the plot the times coincide.}
\label{figthree}
\end{center}
\end{figure}

The last step in the reduction is the summing of the zonal and sectoral flow velocity deviations for each latitude band into temporal bins of 6 Carrington rotations in duration.  For each bin of 6 Carrington rotations we obtain 42 values of zonal flow velocity deviations and 42 values of sectoral flow deviations.  Because the time interval for 6 Carrington rotation is not an even fraction of the year, the x-axis points drift relative to the year.

The flow deviation grids can be plotted as flow maps.  These maps are shown in Figure (\ref{figthree}).  The similarity between the patterns of speed-up and slow-down for these two flows is remarkable.  The white lines were drawn by eye on the zonal flow map to correspond roughly with the location of the zonal flow maxima.  The lines were then transfered to the sectoral flow map to allow comparison of the spatial-temporal patterns.  The combined flow patterns are what constitute the surface pattern of the Torsional Oscillations.  The patterns have greater regularity than has been seen before and sectoral flow extends to higher latitude than shown by \cite{2021ApJ...908L..50G}.

\section{Time Slice Plots - Solar Cycle Dependence}

The 2D color plots give an overall view of the behavior of the zonal and sectoral flows.  However the forms of the functions are not readily apparent.  There are several shifts in $\delta V_{\rm sectoral}$ occurring at different latitudes and showing too many curves on a single plot leads to a spider-web result where the consistent features cannot be seen.  We have previously noted [\cite{2022RNAAS...6..181U}] that $\delta V_{\rm sectoral}$ shows upwelling at sunspot minimum.  Based on this idea we have segregated the time slice plots based on the strength and sign of the high latitude sectoral velocity.  The phase of the solar cycle is well followed with this discriminant.   

\begin{figure}
\begin{center}
\resizebox{4.5in}{!}{\includegraphics{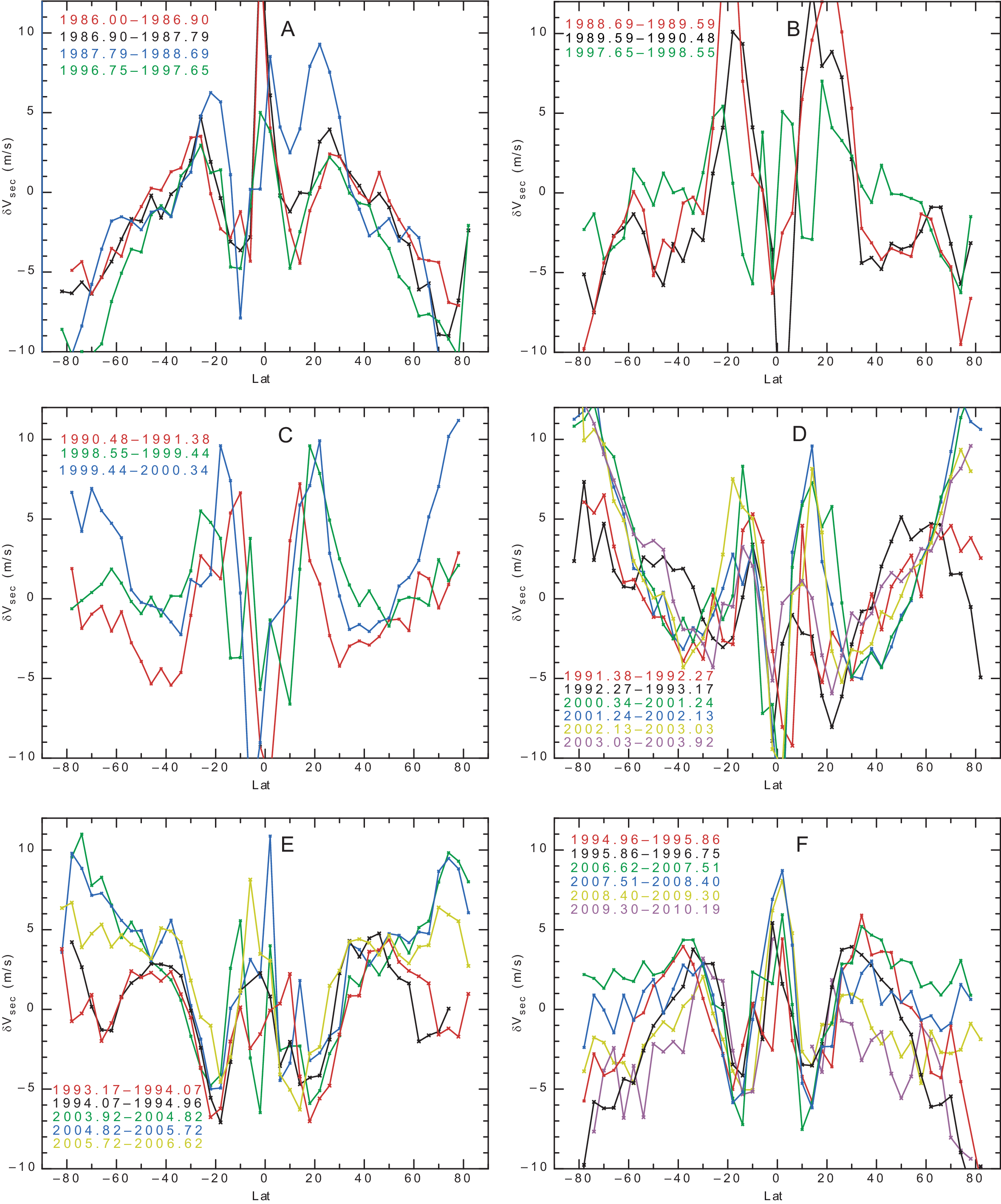}}
\caption{Time slice plots of $\delta V_{\rm sectoral}$ versus Latitude for 12 Carrington Rotation averages.  The curves are color coded and the times of the averages are show on the left with the characters of the times having the same color as the lines.  The times are the start and finish for each temporal average.  The choice of panel for showing each plot was based on both the high latitude flow value and the active latitude flow value.  The spacing started with two lines per panel but then some lines needed to be moved forward or backward to maintain the most nearly similar shapes.}
\label{figfive}
\end{center}
\end{figure}

Figure (\ref{figfive}) shows the time slice plots segregated according to the time within the solar cycle which picks the selection very well according to the high latitude values of $V_{\rm sectoral}$.  Panel {\bf A} is at the times of minimum with the strongest polar outflow.  The active region bands near $15^\circ-20^\circ$ N/S have equatorial direction flows as well.  Panels {\bf B} and {\bf C} are transitioning toward solar maximum with the high latitudes going toward poleward flow and the active region bands also developing stronger poleward flow.  Panel {\bf D} is after solar maximum but still has the active region bands flowing poleward while the high latitudes are still strongly poleward in their flow.  Panels {\bf E} and {\bf F} are transitioning to solar minimum with the high latitude flow becoming equatorward in panel {\bf F}. In both these panels the active latitude bands are equatorward in their flow.  It is worth noting that the time between solar maximum and solar minimum was longer for cycle 23 than for cycle 22 yet the dynamics of the sectoral flow continued to follow the pattern as minimum was approached in cycle 23.

\section{Conclusions}

This study has yielded sectoral flow velocities at latitudes which are much closer to the poles compared to what has been available before.  The pattern seen in Figure (\ref{figfour}) is consistent with expectations for the meridional flow between latitudes of 15$^\circ$ and 60$^\circ$.   Nearer the equator the determination here is uncertain due to the possible vertical flow which is being interpreted by our formulations as the projection of a flow along a level surface.  The strong negative values in Figure (\ref{figfour}) are no doubt the result of a subsidence at the equator.  Near the poles the return flow from the time variable circulation currents dominates the observed sectoral flow and includes periods when the polar upwelling generates strong equatorward flows.  It can be hoped that the new observational results will help the development of a more complete theory of the solar circulation.

The Torsional Oscillations are well studied and it is well known that the migration of the pattern from high latitudes to the equator requires most of two solar cycles.  This study has shown that the sectoral flow deviations and the zonal flow deviations are well synchronized during these cycles.  The Torsional Oscillations evidently involve motions in the N/S direction as well as the E/W direction.  The high latitude polar outflow at sunspot minimum found by \cite{2022RNAAS...6..181U} extends to a polar inflow at sunspot maximum.  The pattern of high latitude flow deviations is dependent on the phase of the sunspot cycle and can provide a new way of predicting the time remaining before the next sunspot minimum.

%
\begin{acks}
The Mt. Wilson Observatory project was started by the Carnegie Institution of Washington and financed by that institution until 1984 when management and funding were transferred to UCLA.  Before the transfer funding was provided by the US Navy through its Office of Naval Research as well as NASA and NSF.
After the transfer NASA and NSF provided most of the support along with continuing support from ONR.
Current support comes from NSF through grant 2000994 while earlier support came from the NSF through grant AGS-0958779 and NASA through grants NNX09AB12G and HMI subcontract 16165880. Over the years additional funding has come from these two agencies as well as the ONR and NOAA..  Ulrich and Boyden are supported by retirement funds while Tran is supported by the NSF grant.  The observatory is managed by the Mt. Wilson Institute.
\end{acks}

\bibliographystyle{klunamed_rku}
\bibliography{rku-lib-1}
\end{article}

\end{document}